\documentclass[aip,twocolumn,letterpaper]{revtex4}

\usepackage{fullpage}

\usepackage{xcolor}
\definecolor{ultramarine}{rgb}{0.07, 0.04, 0.56}

\usepackage{graphics}
\usepackage{epsfig}

\usepackage[margin=1.0in]{geometry}

\begin{document}
\title{Comment on Nuclear Fusion \textbf{66}, 016012 (2026) by Richard Fitzpatrick, \\ \emph{A Simple Model of Current Ramp-Up and Ramp-Down in Tokamaks} }
\author{Allen H Boozer}
\affiliation{Columbia University, New York, NY  10027 \linebreak ahb17@columbia.edu}

\begin{abstract}

The paper Nuclear Fusion \textbf{66}, 016012 (2026) by Richard Fitzpatrick is based on fundamental errors in the physics of the evolution of the poloidal magnetic flux in tokamaks.   With the certainty of Faraday's Law, the slippage of the poloidal relative to the toroidal flux is given by the loop voltage.  Some of the errors may be endemic to the tokamak community, such as the focusing on a quantity $\psi$ that has a subtle relation to the poloidal flux while essentially ignoring the toroidal flux.  Other errors are due to an entanglement of in principle separable concepts and unjustified assumptions.  Although Nuclear Fusion allowed the use ``private communication" as a reference to bolster misquotes, the response must be in arXiv because Nuclear Fusion will not allow any mention of  ``private communication" in a response.  This comment also derives a new and simple expression for the loop voltage, which has the validity of the parallel component of Ohm's law. 

\end{abstract}

\date{\today} 
\maketitle

\section{Introduction}

Richard Fitzpatrick's 2026 Nuclear Fusion paper \cite{Fitzpatrick:2026} on current ramp-up and ramp-down in tokamaks is based on misunderstandings about the form of Faraday's Law in toroidal plasmas, which may be endemic within the tokamak community, coupled with errors due to entangling concepts that in principle separable and making unjustified assumptions.  The scientific issues in his paper are dealt with in Section \ref{Scientific} and even in more detail in my Physics of Plasmas paper \emph{Constraints on the magnetic field evolution in tokamak power plants} \cite{Constraints:2026}.

There is a second set of issues that arose from misquotes in the Nuclear Fusion paper \cite{Fitzpatrick:2026} and the use of references to ``private communication" to bolster those misquotes.  As discussed in Section \ref{Misquotes}, all the ``private communication" was through emails which explicitly pointed out the falseness of his quotes rather than supported them.

The Nuclear Fusion paper \cite{Fitzpatrick:2026}, which was accepted on October 14 and published online of November 6, 2025, was Fitzpatrick's response to a September 9, 2025 arXiv version of my article \cite{Constraints:2025}, which was eventually published by the Physics of Plasmas \cite{Constraints:2026}.  However, the initial submission was flatly rejected on Nov 10, 2025.  The first reviewer of  \cite{Constraints:2025} stated: ``The discussion of the plasma current ramp-down is similarly qualitative, and a more rigorous model was recently published in Nuclear Fusion by Fitzpatrick, prompted by the arXiv article from this author.''  The Physics of Plasmas required that the issues raised in the Nuclear Fusion paper \cite{Fitzpatrick:2026} be addressed before the paper \emph{Constraints on the magnetic field evolution in tokamak power plants} would receive any consideration. 

Without being aware of Fitzpatrick's submission and certainly without permission, his Nuclear Fusion paper \cite{Fitzpatrick:2026} was published with references to ``private communication'' to buttress incorrect statements of the views expressed in the arXiv article \cite{Constraints:2025}.  Despite having copies of all of the ``private communication'' emails,  the journal Nuclear Fusion, as a matter of policy, will not allow publication of a Comment that discloses the actual contents of a ``private communication'' in order to correct the misstatements.  This comment, which will only be published on arXiv, is a companion to a comment written for publication in Nuclear Fusion that only deals with the scientific issues.

%%%%%%%%%%%%%%%%%%%%%%%%%%%

\section{Scientific Issues \label{Scientific}}

\begin{figure}
\centerline{ \includegraphics[width=2.0 in]{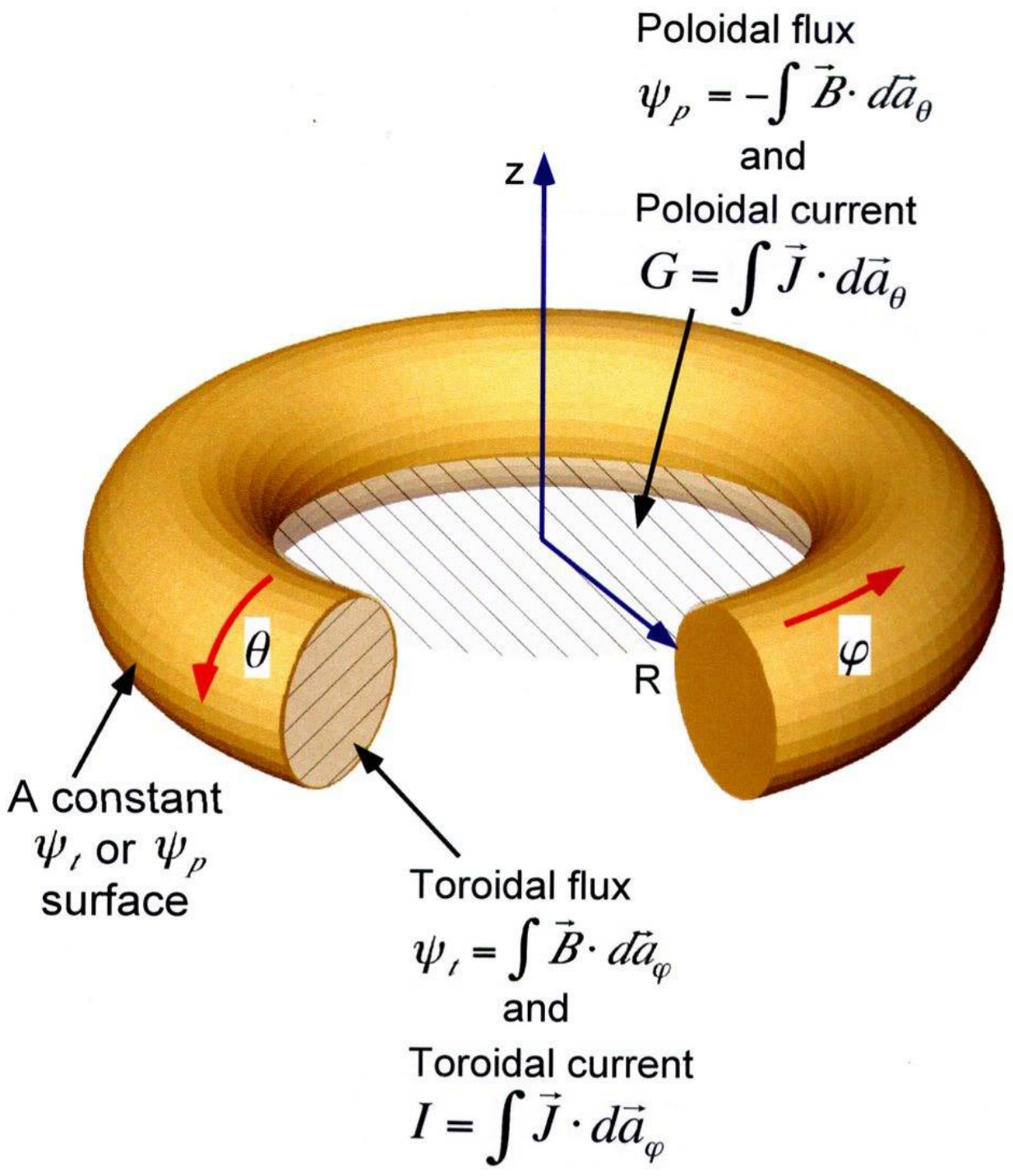}}
\caption{The toroidal flux $\psi_t$ is the magnetic flux enclosed by a toroidal magnetic surface. The poloidal flux $\psi_p(\psi_t,t)$ is  the magnetic flux going down through the hole in the center of the toroidal magnetic surface. The toroidal current is $I(\psi_t,t)$ is the current enclosed by the magnetic surface.  The poloidal current $G(\psi_t,t)$ is the current coming up through the hole in the center of the toroidal magnetic surface.  The current density in the figure is denoted by $\vec{J}$, but in this paper by $\vec{j}$.  This was Figure 1 in Boozer,  Nucl. Fusion \textbf{55}, 025001 (2015).  } 
\label{fig:fluxes-currents}
\end{figure}

Fitzpatrick's Nuclear Fusion paper \cite{Fitzpatrick:2026}  was based on a misunderstanding Faraday's Law.  Judging by the exuberance of the reviewer of \cite{Constraints:2025} and the acceptance and online publication of \cite{Fitzpatrick:2026} by Nuclear Fusion on November 6, 2025, this misunderstanding may be endemic within the tokamak community.  A fundamental problem is the failure to recognize the equation
\begin{eqnarray}
\frac{\partial \psi_p(\psi_t,t)}{\partial t} &=& V_\ell(\psi_t,t),  \label{flux-ev} 
\end{eqnarray}
which has the validity of Faraday's law in toroidal plasmas in which there are magnetic surfaces.  The toroidal flux is the magnetic flux enclosed by a magnetic surface, $\psi_t = \int \vec{B}\cdot d\vec{a}_\varphi$, and the poloidal flux is the magnetic flux that passes down through the central hole in a magnetic surface, $\psi_p= - \int \vec{B}\cdot d\vec{a}_\theta$, Figure \ref{fig:fluxes-currents}.  The loop voltage $V_{\ell}$ gives the slippage of the poloidal relative to the toroidal flux. 

%%%%%%%%%%%%%%%%%%%%%% 
%%%%%%%%%%%% 

The derivation of Equation (\ref{V-I}) in Appendix \ref{App-V}  shows that the Ohm's law for the electric field parallel to the magnetic field, $E_{||}= \eta j_{||} - j_{bs} -j_{cd}$, implies that 
\begin{eqnarray}
 \frac{dI}{d\psi_t} &=&  \frac{dI_{bs}}{d\psi_t} +  \frac{dI_{net}}{d\psi_t} + \frac{V_\ell}{ \mu_0 (G+\iota I) \eta},  \label{I'}
\end{eqnarray}
where $I=\int \vec{j}\cdot d\vec{a}_\varphi$ is the toroidal current enclosed by a magnetic surface, $G=\int \vec{j}\cdot d\vec{a}_\theta$ is the current upward through the central hole in a magnetic surface, Figure \ref{fig:fluxes-currents}, and $\iota=d\psi_p/d\psi_t$ is the rotational transform.  The currents $I_{bs}$ and $I_{cd}$ are the total bootstrap and the total externally driven current enclosed by a magnetic surface.

Equations (\ref{flux-ev}) and (\ref{I'}) are easiest to derive in what are known as Boozer coordinates \cite{Mag-coord} in the stellarator community, Figure \ref{fig:fluxes-currents}:  
\begin{eqnarray}
2\pi \vec{B} &=& \vec{\nabla}\psi_t \times \vec{\nabla}\theta + \vec{\nabla}\varphi \times \vec{\nabla}\psi_p(\psi_t,t). \label{contra-mag-coord} \\
&=& \mu_0 G(\psi_t,t) \vec{\nabla}\varphi + \mu_0 I(\psi_t,t) \vec{\nabla}\theta + \beta_*\vec{\nabla}\psi_t.  \hspace{0.2in} \label{co-mag-coord} 
\end{eqnarray}
Boozer coordinates exist in the magnetic field of any toroidal scalar-pressure plasma equilibrium.  These coordinates are well known among physicists working on stellarators since they are the basis of the optimization that led to the W7-X stellarator and the modern stellarator program.  The proof of Equation (\ref{co-mag-coord}) is non-trivial even in Boozer coordinates but is given in \cite{Boozer:RMP}.
%%%%%%%%%%

Equations (\ref{flux-ev})) and (\ref{I'}) are coordinate independent expressions.  The failure to grasp their importance presumably comes from the rarity of a mention of the toroidal magnetic flux in the tokamak literature.  Instead of using the toroidal flux $\psi_t$ as the radial coordinate, as is standard in the stellarator literature, the tokamak literature uses $\psi$.  The flux $\psi$ is the natural solution to the Grad-Shafranov equation but has a non-trivial relationship to the poloidal flux $\psi_p$,
\begin{equation}
\psi \equiv \Psi_p^{ax}-\psi_p,
\end{equation}
where $\Psi_p^{ax}(t)$ is the magnetic flux passing downward through the region enclosed by the magnetic axis.  A rigorous discussion of the implications of Faraday's Law, the slippage of the poloidal flux $\psi_p$ relative to the toroidal flux $\psi_t$, is made subtle by the use of $\psi$ as the radial coordinate and impossible when the toroidal flux $\psi_t$ is ignored. 

\begin{figure}
\centerline{ \includegraphics[width=1.8 in]{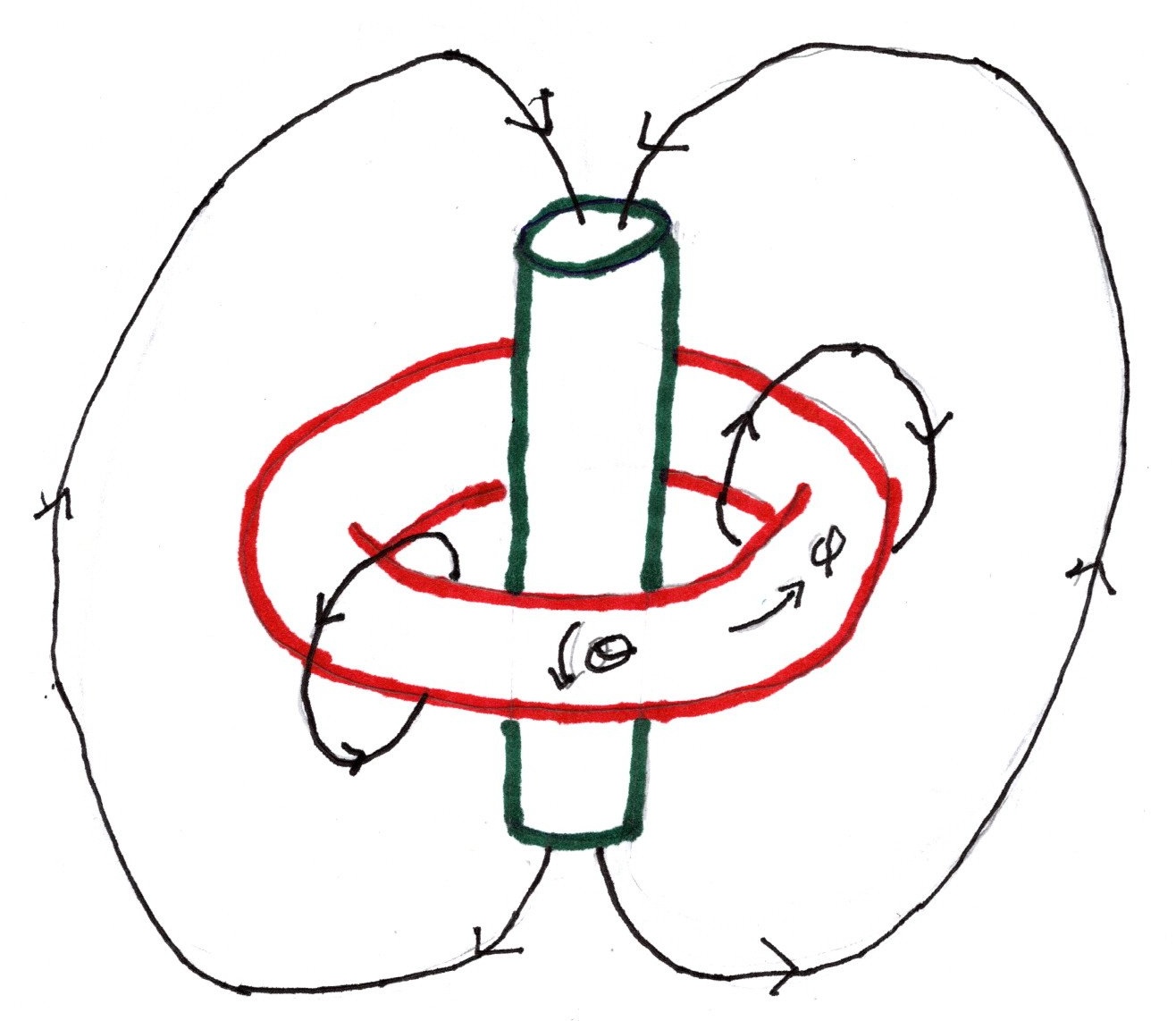}}
\caption{The lines of the poloidal magnetic field produced by the toroidal plasma current are shown together with the magnetic field produced by the central solenoid of a tokamak. }
\label{fig: B}
\end{figure}

Although Equation (\ref{flux-ev}) is non-trivial to prove, it is trivially provable at the magnetic axis, $\psi$=0.  At the magnetic axis, the application of Stokes' Theorem to Faraday's Law proves the time derivative of the downward magnetic flux through the region enclosed by the magnetic axis, $\Psi_p^{ax}$, is given by 
\begin{eqnarray}
\frac{d\Psi_p^{ax}}{dt} = V_\ell^{ax};   \label{Psi_ax}\\ 
V_\ell^{ax}\equiv\oint_{ax}\vec{E}\cdot d\vec{\ell}. 
\end{eqnarray}
Even this trivially obvious expression implies
\begin{eqnarray} 
\Psi_p^{ax} &=& \Psi_p^{sol} + \Psi_p^{pl};  \label{Axis flux}\\
\Psi_p^{pl} &=& - \Big(\Psi_p^{in} + \Psi_p^{ex}\Big), \label{Plasma flux}
\end{eqnarray}
where $\Psi_p^{sol}$ is the magnetic flux coming downward through the central solenoid of a tokamak and $\Psi_p^{pl}$ is the poloidal flux going downward through the region enclosed by the magnetic axis due to the plasma current, Figure \ref{fig: B}.  The poloidal flux due to the plasma current is conventionally negative and is the sum of two parts.  The part $\Psi_p^{in}$ gives the poloidal flux due to the plasma current that lies within the plasma, and $\Psi_p^{ex}$ is the part of the poloidal flux due to the plasma current that lies outside of the plasma.  Normally, $\Psi_p^{ex}$ is somewhat larger than $\Psi_p^{in}$.   The Nuclear Fusion paper \cite{Fitzpatrick:2026}, which was inspired by an arXiv article \cite{Constraints:2025},  would have avoided many of its errors had Figure \ref{fig: B}, which was in the arXiv article, not been ignored.

%%%%%%%%%%%%%%%%%%%%%%
%%%%%%%%%%%%

Richard Fitzpatrick's 2026 Nuclear Fusion paper on current ramp-up and ramp-down in tokamaks \cite{Fitzpatrick:2026} has all the confusions that would be expected from basing the magnetic field evolution on the evolution of $\psi$ and ignoring $\psi_t$.  In addition, the paper entangles a number of in-principle separable concepts and made incorrect assumptions.  The problems with his Nuclear Fusion paper \cite{Fitzpatrick:2026} are dealt with in \cite{Constraints:2026}, and that paper should be consulted for details.  Only a summary of the problems and incorrect assumptions are given here:

\begin{enumerate}

\item Did not mention the changes in the flux in the central solenoid, $\Psi_{sol}(t)$, which is the only flux that is a directly controllable function of time.

\item Assumed the magnetic flux enclosed by the axis is $\Psi_p^{ax}(t)$ can be effectively controlled, but this flux is given by Equation (\ref{Axis flux}) and determined by two terms, Equation (\ref{Plasma flux}), that are not even discussed $\Psi_p^{sol}(t)$ and $\Psi_p^{ex}$, which is most of the poloidal flux produced by the plasma current.

\item Assumed a spatially constant loop voltage in the plasma without explanation, which happens to be true \cite{Constraints:2025}.

\item Assumed a spatially constant loop voltage implies a profile of the magnetic field $B_\theta(r,t)/B(a,t)$ that is independent of time with $a$ the plasma radius in the cylindrical model that was used.  This assumption is false.  When the plasma radius $a$ is fixed, the converse is true.  When $B_\theta(r,t)/B(a,t)$ is independent of time, the loop voltage is spatially constant.  The bootstrap and driven currents were assumed zero.   

\item In the simple cylindrical model of  \cite{Fitzpatrick:2026}, $B_\theta(r,t)/B(a,t)$ and $a$ independent of time would imply the current profile $I(r,t)/I(a,t)$ is independent of time.  But, Equation (\ref{I'}) says the current profile is dependent on the resistivity $\eta$ profile which depends on both the electron temperature and the $Z_{eff}$ profiles.   Even with a fixed impurity content, $Z_{eff}$ can be a complicated function of the electron temperature.  

\item The assumption of a temporally constant current profile meant that if the initial plasma was stable based on the 1987 study of TFTR disruptions by Cheng, Furth, and Boozer \cite{MHD stab}, which was used, the plasma would remain stable. 

\item Assumed a time-independent and spatially-constant profile of the thermal energy diffusivity to calculate the electron temperature and that $\eta j^2$ was the only other term in the power balance.  These assumptions made $T_e(r,t)/T_e(0,t)$ independent of time, but neglected the effects such as the loss of fusion heating and the loss of the H-mode in a realistic plasma model.

 \item Non-spatially constant thermal diffusivities were mentioned, which would have given different $T_e$ and hence $I(r)$ profiles, but only one current profile was considered and that was assumed to be applicable to all current profiles consistent with plasma cooling on the timescale given by the spatially constant thermal diffusivity.   A remarkable extrapolation was made that all ramp downs of similar speed are stable.  Only a limited range of current profiles are stable using the Cheng et al study \cite{MHD stab}.

\item Had no response to the sentence in the abstract of  \cite{Constraints:2025}: ``A deviation of the profile of the plasma current over its full stability range produces only a small change, $\sim16\%$, in the poloidal flux produced by the plasma current. This offers a simple explanation of why disruptions in tokamaks are so common, and why current-profile control though difficult seems to be required, especially during shutdown.''

\end{enumerate}

The Nuclear Fusion paper \cite{Fitzpatrick:2026} is a complicated intertwining of confusions about Faraday's Law,  invalid assumptions,  and incorrect approximations.  The paper considers only one current profile, which appears to be stable, but this is not a credible argument for the robust stability against disruptions for all realistic current profiles.  

Two statements are obviously correct but not reflected in the analysis in the Nuclear Fusion paper \cite{Fitzpatrick:2026}:  (1) The definition of the poloidal flux must include the solenoidal flux in order to represent its induction effect.  (2) The time derivative of the poloidal flux is at a fixed toroidal flux not at a fixed minor radius $r$, as is obvious when the plasma is shifted outward in major radius $R$ assuming an ideal magnetic evolution.  Confusion may arise because these two statements are only implicit in standard tokamak references.   

With 1600 downloads, the errors in \cite{Fitzpatrick:2026} may be sufficiently subtle to confuse many people and not just the usually reliable reviewers for the journals Nuclear Fusion and Physics of Plasmas.

Whatever the correctness of \cite{Fitzpatrick:2026} may be, the reviewers may have been lulled into acceptance from knowledge of results based on codes that study magnetic induction in tokamaks. These codes have undergone decades of development and testing against thousands of tokamaks discharges on an extremely large range of device sizes, plasma temperatures, and engineering designs.  But, this seems unlikely.

If  these codes were adequate for designing tokamaks that are sufficiently disruption free for practical fusion power, it should be possible to determine how JET-ILW should have been operated to have high-performance disruption-free plasmas.  This would  show the naivete of the concerns of Eidietis \cite{Eiditis:2021} and of the statement in \cite{JET-dis:2020}: ``It may be expected that a disruption free space may be defined in the $\ell_i - q_{95}$ empirical stability diagram, assuming that plasma current profiles tend to maintain itself inside the permissible values. In reality, the JET-ILW pre-disruptive plasma equilibrium parameters create a diffused cloud on the $\ell_i - q_{95}$ stability diagram without room for non-disruptive plasmas.''
 
 An easier task for the codes would be the identification of parameters that can be monitored and controlled in a fusion power plant that would ensure a sufficiently low disruption rate.  Even that seems too ambitious.  The paper \emph{ARC disruption physics and strategy} \cite{Sweeney:2026}  displays little confidence in the use of existing codes to ensure disruption avoidance and focused on disruption mitigation instead.  The inexplicability of the acceptance of the results of \cite{Fitzpatrick:2026} raises concerns about the reliability of other assessments.

%%%%%%%%%%%%%%%%%%%%%%%%%%%%%%%%%%%%%%%%%%%%%%%%%%%%%%%
%%%%%%%%%%%%%%%%%%%%%%%%%%%%%%%%%%%%%%%%%%%%%%%%%%%%%%%%

\section{Misquoting Issues \label{Misquotes} }

The scientific issues with the Nuclear Fusion article \cite{Fitzpatrick:2026} are now adequately dealt with in my Physics of Plasmas paper \cite{Constraints:2026}.  What is not dealt with are the false statements about my arXiv article that were buttressed by ``private communication'' references in the Nuclear fusion paper \cite{Fitzpatrick:2026}.  The Nuclear Fusion policy of allowing false statements to be buttressed by ``private communication'' references but not allowing any mention of a ``private communication'' to refute those false statements left this arXiv publication as the only channel for refuting them.

 If Fitzpatrick had just ignored the ``private communication" emails, that would have been irresponsible. But, Richard coupled his references to my arXiv article and his references to ``private communication'' everywhere except in the Abstract to reinforce his claim to be correcting a false view of mine.
 
 A particularly clear example of a misquote that went against an email exchange was in the abstract to the Nuclear Fusion paper \cite{Fitzpatrick:2026}:  ``Hence, there is no indication that the design ramp times are infeasible, as was recently suggested in Boozer \cite{Constraints:2026}.''  My August 13, 2025 email, which was before August 18 when the the Nuclear Fusion paper \cite{Fitzpatrick:2026}  was submitted:  ``Actually I didn't really mean the expected ramp down time was 1000 sec, and it would take a lot of power to keep the plasma going. What I was trying to indicate is that a large cooling of the plasma is necessary but the profile had to remain relatively fixed while doing that ramp down. I agree my words were confusing, and they have been changed."  Fitzpatrick responded: ``Would it be accurate to say that you think that the ramp-down times are longer than the SPARC and ITER people are claiming"   My response: ``The primary issues are the energy and magnetic flux input into the plasma that is required during a shutdown to ensure the current profile does not slip into a disruptive state more than one time in at least a thousand shut-downs.  A major problem in my view, is that a 20\% change in the poloidal flux distribution is sufficient to take a plasma far from a state as far as possible from a disruption into a disruption. This means a disruption can occur in approximately a fifth of the time one would expect a flattop to last.''

Despite my statement, ``I agree my words were confusing, and they have been changed," which was an effort to be polite, the  basis of Fitzpatrck's confusion is unclear.  Two quotes  from the first arXiv version my article stated:  (1) ``In principle, the restart period can be made arbitrarily short by clever choices of the time dependence of $\Psi_{sol}(t)$ and the heating power. Of course, the periods in which substantial heating power is required must be short compared to the periods of fusion burn to have net energy to sell.''  (2) "A rapid cooling of the central part of the plasma seems required, but unless this is done with care, the current profile will evolve into a disruptive state.''

The last ``private communication" that Fitzpatrick and I had before the publication of his paper was on 9-10 September 2025, which was before he submitted his revised version of his article, 25 September 2025, the version that was accepted.  This communication illustrates our different perceptions of physics issues.  I wrote on September 9: ``The confusions on the important points in my arXiv article `Constraints on the magnetic field evolution in tokamak power plants' were convincing that the abstract and introduction had to be rewritten. Any comments are appreciated, especially ones on balance and fair quoting.''  Fitzpatrick's September 10 response:  ``I do not really agree that tokamak plasma are ``self-organized''. When I think of the term self-organized, I naturally think of an RFP plasma. Such plasmas are subject to continuous large-amplitude m=1 and m=0 tearing modes. It is these modes that relax the plasma toward a Taylor state. However, a tokamak plasma is quiescent. There are no large amplitude tearing modes, and the equilibrium profiles are determined by transport rather than MHD activity. The only exception is the sawtooth oscillation, which periodically generates a short burst of MHD activity that prevents the central q value from falling significantly below unity. Given that q(0) is pinned to a value close to unity, and q(a) is fixed by the plasma current, the average broadness of the current profile is easily controlled. Experimentalists have also become adept at further tailoring the current profile by means of off-axis heating. It is not true that there is an inevitable drift to a disruptive state in a tokamak plasma. In fact, tokamak discharges could be maintained indefinitely were it not for the limit in the flux-swing of the central solenoid, and the gradual build up of impurities."  

Fitzpatrick's response reads as if disruptions can be easily avoided by a competent machine operator.  Ensuring tokamaks have a sufficiently low disruption rate for the economic feasibility of power plants is non-trivial and requires improved understanding \cite{Eiditis:2021}.  Ensuring disruptions are sufficiently rare for the feasibility of tokamak power plants would seem to require the identification of the externally controllable plasma parameters that show the plasma is safe against disruptions.   This requirement is more stringent in power plants than in existing tokamaks or in JET because of the limitations on diagnostics and actuators.  Generative AI using data from both experiments and simulations could help in this identification.

%%%%%%%%%%%%%%%%%%%%%%
%%%%%%%%%%%%%%%%%%%%%%%%%%%%%%%%%%%%%%%%%%%%%%%%%%%%%%%%%%

%%%%%%%%%%%%%%%%%%%%%%%%%%%%%%%%%%%%%%%%%
%%%%%%%%%%%%%%%%%%%%%%%%%%%%%%%%%%%%%%%%%

\appendix
\vspace{0.2in}
\section{The Loop Voltage \label{App-V}}

This appendix will derive a new and simple expression for the loop voltage, which has the generality of the parallel Ohm's Law.  The definition of the loop voltage, which has the certainty of Faraday's Law when applied to toroidal plasmas, is \cite{Boozer:RMP}
\begin{eqnarray}
V_\ell(\psi_t,t) &\equiv& \lim_{L\rightarrow\infty} \frac{\int_{0}^L \vec{E}\cdot d\vec{\ell}}{\int_{0}^L \vec{\nabla}\left(\frac{\varphi}{2\pi}\right)\cdot d\vec{\ell}}; \label{Loop V},
\end{eqnarray}
where $d\vec{\ell}$ is the  differential distance along a magnetic field line.    Using Ohm's law, the electric field parallel to the magnetic field is $E_{||}=\eta\Big(j_{||} - j_{bs} - j_{cd}\Big)$.  Both the bootstrap current density $\vec{j}_{bs} =j_{bs}\vec{B}/B$ and the current density due to direct current drive $\vec{j}_{cd} =j_{cd}\vec{B}/B$ are defined to be divergence free.

Following a discussion in a paper on the Pfirsch-Schl\"uter current \cite{Boozer:P-S}, the contravariant and covariant representations of the magnetic field in Boozer coordinates, Equations (\ref{contra-mag-coord} ) and (\ref{co-mag-coord} ), can be rewritten at an arbitrary point in time as
\begin{eqnarray}
2\pi\vec{B} &=& \vec{\nabla}\psi_t \times \vec{\nabla}\theta_0;\\
&=& \mu_0(G+\iota I) \vec{\nabla}\varphi + \mu_0 I \vec{\nabla}\theta_0 + B_{\psi_t}\vec{\nabla}\psi_t; \label{new-cov}  \hspace{0.2in}\\
\iota&\equiv& \frac{d\psi_p}{d\psi_t};\\
\theta_0 &=& \theta - \iota \varphi.
\end{eqnarray}
Since  $d\vec{\ell}$ is the differential distance along a magnetic field line
\begin{eqnarray}
\frac{\partial \vec{x}(\psi_t,\theta_0,\ell)}{\partial \ell} &=& \frac{\vec{B}}{B} \mbox{    and  }  \\
\vec{B}\cdot \frac{\partial \vec{x}}{\partial \ell} &=& B,  \mbox{ and using Eq. (\ref{new-cov}) } \\
&=& \Big(\mu_0(G+\iota I) \Big) \Big(\frac{\partial\varphi}{\partial \ell}\Big)_{\psi_t \theta_0}. \hspace{0.2in}
\end{eqnarray}
The implication is that the loop voltage is 
\begin{eqnarray}
V_\ell&=& \mu_0 (G+\iota I) \eta \Big( \frac{dI}{d\psi_t} - \frac{dI_{bs}}{d\psi_t} - \frac{dI_{cd}}{d\psi_t} \Big).  \label{V-I}
\end{eqnarray}

The parallel current $\vec{j}_{||} \equiv j_{||}\vec{B}/B$ is not divergence free, $j_{||} =  j_{ps}+j_{net}$.  The Pfirsch-Schl\"uter current density has a divergence that balances the divergence of the perpendicular  current density that is determined by the pressure gradient, $\vec{j}_\bot\times\vec{B} = \vec{\nabla} p$.  The  Pfirsch-Schl\"uter current  is defined to obey the constraint
\begin{eqnarray}
\lim_{N\rightarrow\infty} \frac{\int_{0}^{2\pi N} \frac{j_{ps}}{B} d\varphi}{\int_{0}^{2\pi N}  d\varphi}=0.
\end{eqnarray}
The net plasma current density is by definiion divergence free, which implies $j_{net}/B$ is constant along a field line, and is given by $j_{net}/B = d I/d\psi_t$, where $I(\psi_t)$ is the toroidal current enclosed by a magnetic surface, $I\equiv\int \vec{j}\cdot d\vec{a}_{\varphi}$, Figure \ref{fig:fluxes-currents}.

%%%%%%%%%%%%%%%%%

%%%%%%%%%%%%%%%%%%%%%%%%%%%%%%%%%%%%%%%%%%%%%%%%%%%%%%%

 %%%%%%%%%%%%%%%%%%%%%%%%%%%%%%%%%%%%%%%%%%%%%

\end{document}